# Study of Gesture Recognition methods and augmented reality

Amol Palve ,Sandeep Vasave

Department of Computer Engineering

MIT COLLEGE OF ENGINEERING

( amolpalve181191@gmail.com ,sandeepvasave17@gmail.com)

*Abstract:*

With the growing technology, we humans always need something that stands out from the other thing. Gestures are most desirable source to Communicate with the Machines. Human Computer Interaction finds its importance when it comes to working with the Human gestures to control the computer applications. Usually we control the applications using mouse, keyboard, laser pointers etc. but , with the recent advent in the technology it has even left behind their usage by introducing more efficient techniques to control applications. There are many Gesture Recognition techniques that have been implemented using image processing in the past.

However recognizing the gestures in the noisy background has always been a difficult task to achieve. In the proposed system, we are going to use one such technique called Augmentation in Image processing to control Media Player. We will recognize Gestures using which we are going to control the operations on Media player. Augmentation usually is one step ahead when it comes to virtual reality. It has no restrictions on the background. Moreover it also doesn't rely on certain things like gloves, color pointers etc. for recognizing the gesture. This system mainly appeals to those users who always looks out for a better option that makes their interaction with computer more simpler or easier.

I.INTRODUCTION:

In Human Computer interaction, the first and foremost thing is the interfacing. There are many techniques that have been implemented so far as interfacing techniques. Gesture Plays an important role while interacting with the Computer applications. It enables humans to communicate with the machine and interact without any mechanical device. In Gesture Recognition , there has been always an issue about how to interact with the Computer. Various bodily gestures are used to interact with  Computer, they include, hands , face, legs, eyes etc but most commonly  used are hands and face. Hand gestures can be used as various commands to operate certain applications. But main problem in Hand gestures is they totally depends on the database that is the storage space for their gesture matching. They are static gestures that needs memory to store the images and learn a particular gesture. For tackling this one of the  problem, Augmentation, was introduced as Human Computer Interaction technology, that aims to combine the 2D or 3D virtual objects with the real world pictures[1].  Augmentation also called as Augmented Reality or Virtual Reality, is more user friendly.  So far we were very used too on the devices like keyboard, mouse etc. This is one such technique that eliminate their usage. The proposed paper focuses  on Gesture Recognition using Augmentation in controlling the media player operations. Media player always requires mouse or keyboard to control it. Every media player application when its active has the basic operations on it. It has also got some shortcut keys that requires the keyboard, to perform some action related with every key. But when the keyboard malfunctions or not working these things become somewhat difficult to control. In this case using  Gestures one can handle any application. Augmented reality  makes  it very user friendly to control any application using gestures.

II.Existing system:

Gesture  Recognition  aims  to  interpret human  gestures  via  Mathematical Algorithms.  There have been many gesture recognition techniques    that are presently worked on such as Color pointer techniques many more.

A.Hand gesture

Modeling of hand gesture mainly require gesture regniation which provide new way ta access  machine[4].for  modeling  mainly consider  two  methods  which  famously knows as spatial and temporal method .this two  method  used  in  different  situation spatial used where shape is important factor while temporal used where motion of hand is consider[4].in 2D and 3D Implementation of  hand  modeling  spatial  is  important one[5].

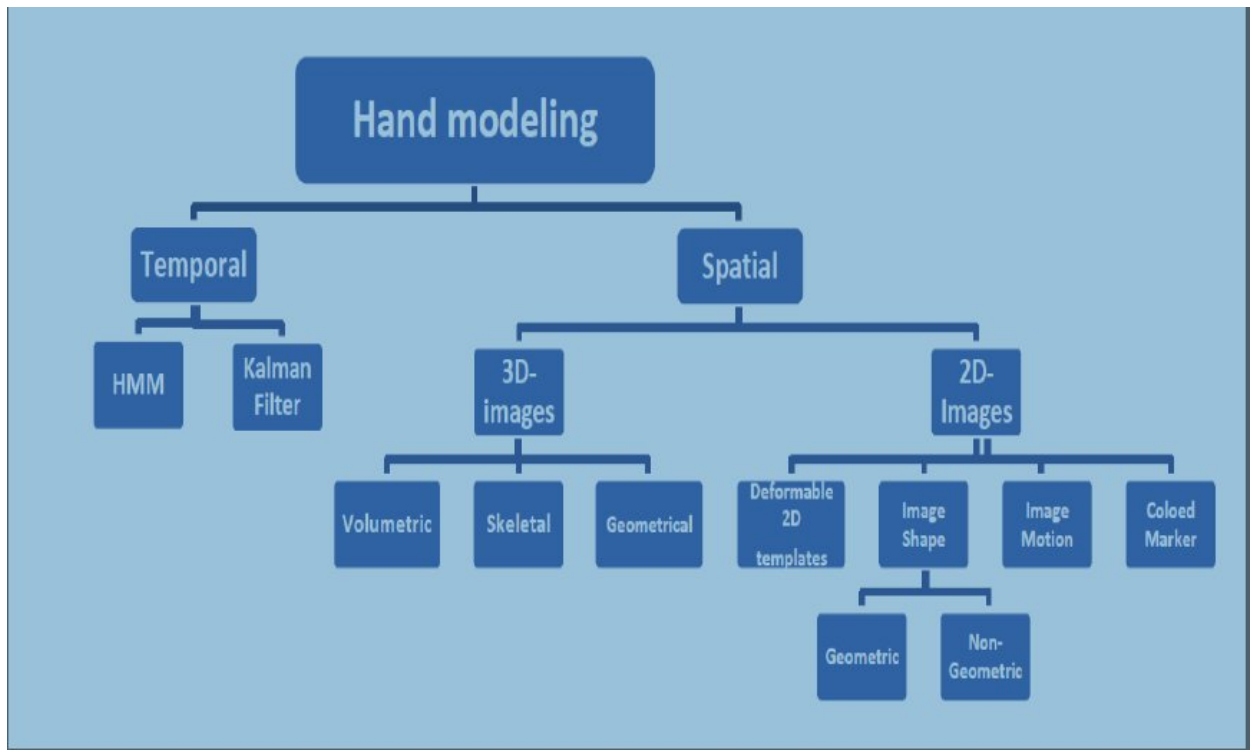

Fig no1.Hand modeling method

Shape having two types geometric and non geometric[6].geometric mainly related to the motion and finger postion and palm properties.while non-geometric consider color and other facters.there are many methods given below which is helpful for hand modeling system

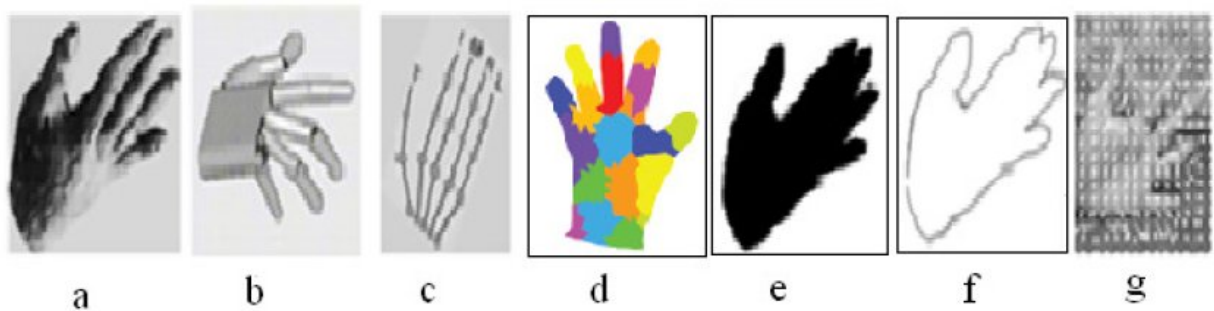

Fig no 2.Methods

Various hand modeling methods to represent hand posture [7]. (a) 3D volumetric model. (b) 3Dgeometric model. (c) 3D Skeleton model. (d) Colored marker based model. (e) Non-geometric shape model(Binary silhouette [5]). (f) 2D deformable template model . Motion based model.

We study different techniques color pointer technique which involves using various color pointers. These color pointers are used on the fingers for Hand gesture recognition[8]. A very popular example of this has been covered by the Pranav Mistry on the Sixth Sense. Another very popular system that exists is the wearable glove that is used for tracking fingers and hand which is a Computer vision based recognition. There exist the optic light based technology which does not usually require user to interact with device by physically touching anything, simple hand motions and gestures can be used to interact with the device[2]. The color models that were used so far was totally based on the RGB .but all methods having some limit on like data glove in which it is compulsory to wear data glove every time and also which are more expensive one[8].while color pointer require that to wear color pointer on hand and if same color appear within regnition area the it may confuse to select object[8].also there are skeleton based method where noise hand diction is very difficult one task[8].also all this method affecting with lighting condition [8].and also all are static in nature where base image required. scale problem come where different size of hand comes in picture .back ground is also important one[9]. so we have to come up with new idea which considerably remove all of this drawback in that case augmented reality is better solution.

### III.Augmented reality

Augmented reality is combination of real world with virtual environment .where it is used in many areas like medical and other areas. in gaming system it play important role[10]. All the draw back which face previously can remove here basic system architecture [10].

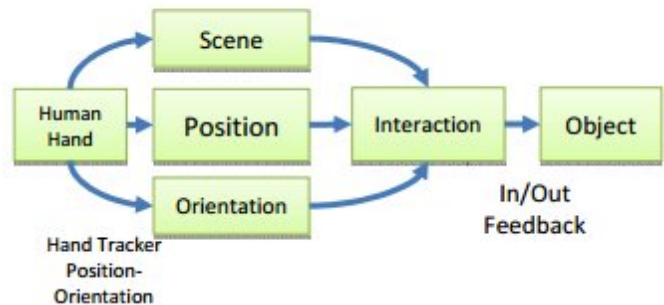

Fig no3.Architeture

IV. Proposed System for hand detection:

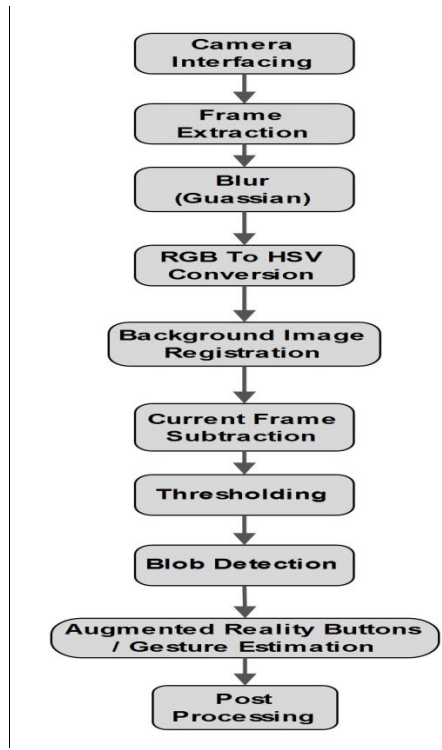

Fig no4. system architecture

In the Proposed system, the main focus is on Recognizing Dynamic Gestures of the user via a Webcam which could be built-in or externally used Camera. Analyzing the gestures and applying the Image Processing algorithms like RBG to HSV conversion, blurring, thresholding, blob detection etc[1].

Algorithms-

A. Bluring image-

After capture image we have to blur image so sharpness reduce so that we get accurete dectcion.

B. RGB TO HSV –

RGB gives more specific value related to image where we require saturated value related that image so we convert into HSV.

C. Thresholding -

Where segmentation can be done easily with the help of thresholding.

D. Blob detecting –

which mainly used to find out particular area from the image.

V. Application-

Using Augmented Reality the gesture will be detected for carrying out certain operations of the application. Virtual menu will be displayed on the live captured feed for action selection. User can directly point towards the button/action for dynamic input. User can also point with the bare hands or using glove or any other object as well. There will be no restrictions on the background. So Augmentation has the advantage for its efficiency.

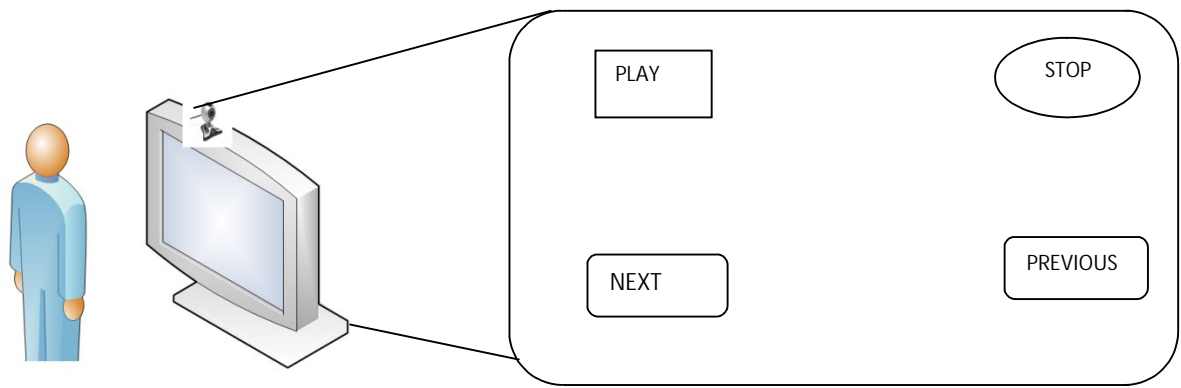

Fig no5.augmented system

In the following diagram, there is a virtual menu displayed on the screen , the User gesture will be feeder through the Webcam and based on the position of the user hands, the gesture will analyzed for the particular Media Player function to be performed. Media Player has to be a active application while using this system.

VI. Conclusion:

This paper presents the system can handle with a Augmented Reality technique. This also focuses on a more User friendly interface between the user and the computer. User can control any applications in the near future using Augmented Reality. User can directly point to the screen and feed the gestures to control the various functions of the Media Player. There is more accuracy rate in the Augmented Reality. Augmentation surely is most promising Human Computer Interaction Technologies to come across. It has a great potential in the coming future to mark a better efficiency in the e-commerce systems as well.

This system has the advantage over other system that, it does not have restrictions on the background. There Dynamic Gestures through live video feed and directly applying the image processing to give output. Also the color model that is used is very strong. Gesture Recognition as a user interface also applies to a wide range of consumer electronics using Augmentation.

is no need of the databases as it is totally works on dynamic gestures behaviour.